# Depressions at the surface of an elastic spherical shell submitted to external pressure.

C. Quilliet

*Laboratoire de Spectrométrie Physique,*
*CNRS UMR 5588 & Université Joseph Fourier,*
*140 avenue de la Physique, 38402 Saint-Martin d'Hères Cedex, France.*
*(permanent address)*
*&*
*Soft Condensed Matter, Debye Institute, Utrecht University,*
*Princetonplein 5, 3584 CC Utrecht, The Netherlands*

**Abstract:** Elasticity theory calculations predict the number N of depressions that appear at the surface of a spherical thin shell submitted to an external isotropic pressure. Using a model that mainly considers curvature deformations, we show that N depends on the relative volume variation and on an adimensional parameter that takes into account both the relative spontaneous curvature and the relative thickness of the shell. Equilibrium configurations show single depression (N=1) for small volume variations, then N increases, at maximum up to 6, before decreasing more abruptly due to steric constraints, down to N=1 again for maximal volume variations. These static predictions are consistent with previously published experimental observations.

## 1. Introduction

Buckling, or sudden change of the shape of objects under constraint, has been a problem addressed for a long time, as it is of utmost practical importance (architecture, designing of tough containers or pipes etc)[1-3]. More recently, buckling was investigated at much smaller scales, mainly due to the improvement of observation and manipulation techniques at microscopic, or even nanoscopic level[4]. This brings back into the spotlight the simplest symmetry that provides a closed vessel, the spherical one. In spite of important theoretical advances motivated by air and spatial navigation[5], the spherical symmetry had been somewhat neglected compared to other symmetries (axisymmetrical, cylindrical etc) because less directly encountered in large scale systems where gravity plays a role. At smaller scales, when gravity becomes negligible, whole spheres are more commonplace: they are encountered in colloids, biological objects, and in most of the phenomena linked to surface tension. Flows, encapsulation processes, evaporation/dissolution phenomena in complex fluids may lead to buckling processes, generating non-trivial shapes. Motivated by recent experimental works[6-8], we focus in this paper on thin spherical shells of an elastic and homogeneous material submitted to an isotropic constraint, such as an external pressure. The onset of buckling of hollow spheres submitted to an external pressure has been derived in earlier times[9,10], recent theoretical work dealing mainly with the non-ideality of materials[11]. For what concerns post-buckling, no analytical predictions concerning post-buckling shapes of spheres under isotropic external pressure are available, by lack of a complete buckling analysis.

Indeed, there is a lively field in biophysics aimed at understanding shapes of biological objects with simple physical ingredients. Since early work by Helfrich and others where bending elasticity and spontaneous curvature[12] (or area-difference elasticity[13], which can be modelized in the same way[14]) were the main shape tuners[15], some models now include additional stretching elasticity, which brings them quite close to what is discussed hereafter. To our knowledge, these are nevertheless inaccurate to interpret the buckling of homogeneous shells, as they precisely take into account inhomogeneities characteristic of the biological objects they aim to describe. For the closest models, these inhomogeneities are either the different origin (lipidic membrane and cytoskeleton) of bending and stretching elasticity[14,16], which allows bending/stretching ratios forbidden in a model of thin homogeneous plate (cf §2), or defects due to the discrete protein structure of a shell virus[17].

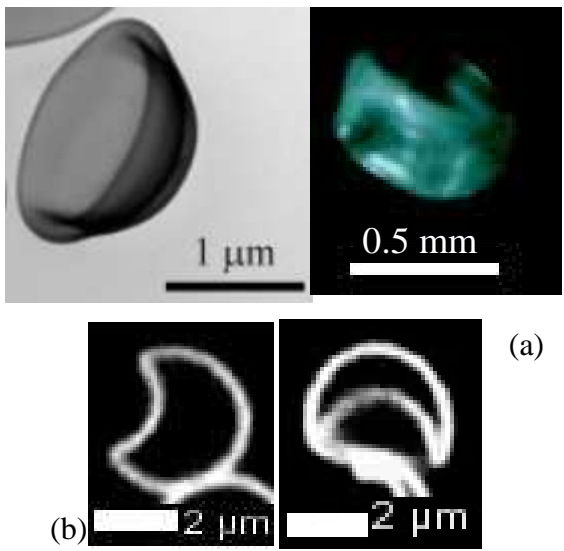

*Figure 1: Different shapes obtained after evaporation of the solvent contained in a spherical porous shell. (a) Left: silica/silicon « capsule » (N=1 ; d/R=0.17) observed by Zoldesi et al[6] (reproduced with author and editor permission). Right: more polyhedral shape (N~6) observed by Tsapis et al in shells made from aggregation of colloïdal particles at the surface of an evaporating droplet of colloidal suspension[7]; d/R=0.09 (reproduced with permission by the author). (b) Deformation of polyelectrolyte capsules (d/R = 0.026) in solution through osmotic pressure (left: 22 kPa ; right: 44.5 kPa)[8].*

In this paper concerning homogeneous objects, we propose a heuristic calculation concerning the optimum number of depressions that can be hold by a buckled thin spherical shell, in order to understand the apparently divergent observations of references [6] and [7]. The two papers reported the observation of strongly buckled objects, originally porous hollow spherical shells filled with a solvent, that buckle when the solvent evaporates. Buckling of a porous shell due to evaporation of an inner solvent is acknowledged as being of a capillary origin, and macroscopically (*i.e.* at scales larger than the pore size) equivalent to the effect of an isotropic external pressure[18]. These experiments therefore constitute a direct illustration of the postbuckling of hollow spheres under external pressure, for which no theoretical predictions exist. Surprisingly, conformations taken by the shells qualitatively differ between the two references [6] and [7]: as shown on Fig. 1a, there is occurrence of either a single and quite deep depression[6], or several depressions distributed over the sphere's surface, leading to a coarsely cubic shape[7]. Single depression was also obtained by osmotic pressure (fig. 1b)[8]. In this paper we will determine whether this discrepancy can be interpreted through the equilibrium configurations of an elastic model, or if some drying artefacts should be invoked.

## 2. Elasticity

We will use elasticity theory results related to thin shells submitted to external constraints in order to get insight into the number of depressions expected for an equilibrium conformation. For a thin spherical shell of homogeneous elastic material (Young modulus $E_{3D}$, sphere radius R, thickness d), deformation occurs by bidimensional stretching (stretch modulus $E_{2D} = E_{3D} d$) or bending (bending modulus $\kappa \sim E_{3D} d^3$). The latter is related to the mean curvature, and we do not take into account energy variations linked to the Gaussian curvature, as its integral on a closed surface depends only on the topology (Gauss-Bonnet theorem). It has been shown[9] that the ratio between bending and bidimensional stretch energies scales as :

$$U_b/U_s \sim (d/R)^2$$

where R is the radius of the spherical shell, and d its thickness. Hence for sufficiently thin shells, deformations occur mainly by bending, which is much less energetic than stretching. In this framework, a depression obviously corresponds to the inversion of a spherical cap[9,10], which avoids stretching energy outside the circular ridge that joins the undeformed part and the inverted cap, defined by its half-angle α (Fig. 2). For shells of zero spontaneous curvature (the case with nonzero spontaneous curvature will be treated in the paragraph 3), bending energy does not depend on the sign of the curvature and hence is not modified in the inverted cap.

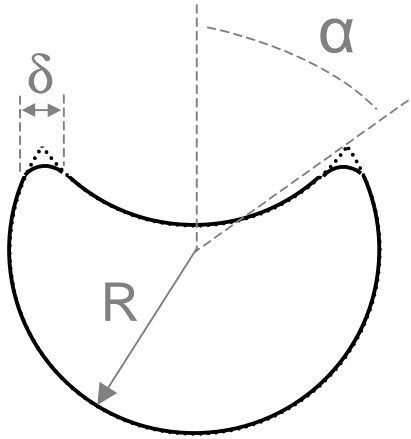

*Figure 2: Depression formed by inversion of a spherical cap. The circular ridge that allows a continuous jonction between the undeformed spherical part and the inverted cap has a lateral extension $\delta \sim (Rd)^{1/2}$ (where d is the thickness of the shell). The aperture of the depression is defined by the half-angle α that extrapolates (dotted line) the ridge thickness down to zero.*

Energy modifications then concentrate in the circular ridge, which lateral extension is imposed by minimization as $\delta \sim (Rd)^{1/2}$; the relevant curvature radius being $\delta/\tan\alpha$ [10,19]. As the total surface bended in the ridge has a width δ and a length $2\pi R \sin\alpha$, the energy of a single depression writes:

$$U_1 = 2\pi \kappa (d/R)^{-1/2} \sin\alpha \, tg^2\alpha \quad (1)$$

This expression diverges when α approaches $\pi/2$: thin shell theory breaks down when the radius of curvature becomes of order d. However, as will appear obvious later, other considerations different from energetic ones prevail in this limit, and looking for a more accurate expression is unnecessary within the current work.

The volume variation ΔV due to cap inversion is twice the volume of the spherical cap, hence the volume variation relative to the undeformed sphere volume:

$$(\Delta V/V_{sphere})_{1\ depression} = (1 - \cos\alpha)^2 (2 + \cos\alpha) / 2 \quad (2)$$

In the case of N similar depressions, we have:

$$U_N = N U_1 \quad (3a)$$

and

$$\Delta V/V_{sphere} = N (\Delta V/V_{sphere})_{1\ depression} \quad (3b)$$

We therefore have the expressions of both the elastic energy and the volume variation corresponding to N similar depressions formed by inversion of spherical caps of half-angle α. As the relative volume variation is the key parameter to appreciate the deformation intensity, it would be interesting to eliminate α in order to get the elastic energy $U_N$ as a fonction of $\Delta V/V_{sphere}$, and then to discuss the relative stability of the conformations with different numbers N of depressions ("states").

Explicit expression for small depressions:
For small inverted caps ( α << 1), system (3) simplifies to:

$$U_N \sim 2\pi N \kappa (d/R)^{-1/2} \alpha^3 \quad \text{and}$$
$$\Delta V/V_{sphere} \sim N \alpha^4 / 8$$

Which, for a given relative volume variation $\Delta V/V_{sphere}$ leads to:

$$U_N \sim 8^{13/12} \pi N^{1/4} \kappa (d/R)^{-1/2} (\Delta V/V_{sphere})^{3/4} \quad (4)$$

For a given shell (E, d and R fixed), the $N^{1/4}$ dependence with $\Delta V/V_{sphere}$ clearly leads to N=1 at equilibrium. The conformation with a single depression is favored in this regime, which corresponds to small deformations ($\Delta V/V_{sphere} << 1$).

Implicit dependence for larger depressions:
For larger values of α, a parametric plot of $U_N$ as a fonction of $\Delta V/V_{sphere}$ shows that for increasing $\Delta V/V_{sphere}$, the lowest energy state changes from N=1 to 2 to 3 etc (Fig. 3). However, system (3) is not sufficient to describe the conformations adopted at high deformations, since equation (3b) is valid only as long as depressions do not interpenetrate each other. There is of course a maximum size (*i.e.* maximum α) above which it becomes impossible to find an arrangement of N similar depressions that avoids interpenetration between two of them. For the sake of simplicity, we will neglect the thickness of the circular ridge, which corresponds to the case d<<R. Then the problem becomes purely geometrical: how many similar spherical caps can be inverted without interpenetration at the surface of a sphere?

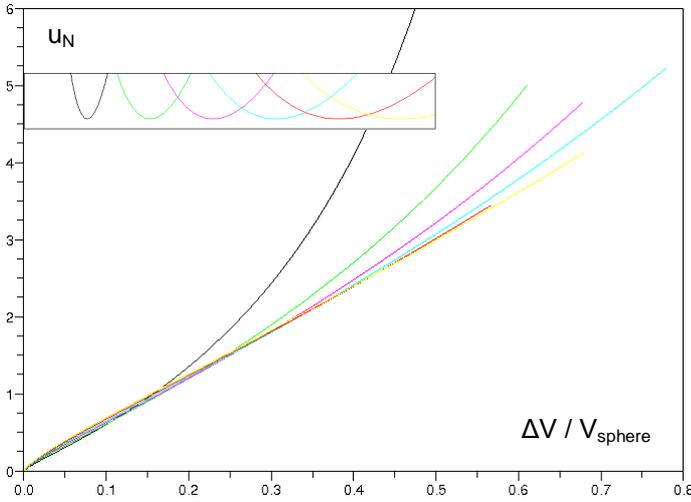

*Figure 3:* Reduced energy $u_N = U_N / (2\pi \kappa (d/R)^{-1/2})$ as a function of $\Delta V/V_{sphere}$, for different numbers N of similar depressions (inverted spherical caps). The curves are traced up to the maximum value $(\Delta V/V_{sphere})_{max,N}$ above which there would be an interpenetration between depressions. Black, green, pink, blue, red and yellow curves correspond respectively from N=1 up to 6. Note the $(\Delta V/V_{sphere})^{3/4}$ behavior of eq. (4) for small $\Delta V/V_{sphere}$, that corresponds to small values of α. Insert: $u_N/(\Delta V/V_{sphere})$ as a function of $\Delta V/V_{sphere}$ (same scale in abscissa), displayed to enhance the difference between curves without modifying the $\Delta V/V_{sphere}$ cross-over values.

A necessary condition is that the spherical caps themselves do not overlap before inversion. This amounts to the search for the maximum surface of a sphere that can be covered with N similar spherical caps, which is another formulation of the so-called Tammes' problem: maximize the minimum point-to-point distance for a set of N points placed on a sphere[20]. The solution of the Tammes' problem is much less obvious than its formulation;

depending on N, analytical or numerical solutions exist[21-23], that for our purpose allow to calculate $(\Delta V/V_{sphere})_{max}$ for $N \geq 5$ (Table 1).

| N | α max (deg) | maximum $\Delta V/V_{sphere}$ |
|---|---|---|
| 1 | 90 | 1 |
| 2 | 60 | 0.625 |
| 3 | 54.7 | 0.691 |
| 4 | 52.2 | 0.785 |
| 5 | 45 | 0.581 |
| 6 | 45 | 0.697 |
| 7 | 38.9 | 0.480 |
| 8 | 37.4 | 0.474 |
| 9 | 35.3 | 0.427 |
| 10 | 33.1 | 0.373 |
| 11 | 31.7 | 0.350 |
| 12 | 31.7 | 0.382 |
| 13 | 28.6 | 0.277 |
| 14 | 27.8 | 0.270 |
| 15 | 26.8 | 0.251 |

*Table I:* Maximum values $\alpha_{N,max}$ of the half-angle α that defines a spherical cap at the surface of a sphere, over which one cannot invert N similar caps without interpenetration. Values of $\alpha_{max}$ for cases N=1 to 4 are specifically treated in the text. Values for N=5 to 8 correspond to the Tammes' problem and are given in reference[22], N=9 in reference[27], N=10 to 15 in reference[23]. For N higher than the values displayed in this table, the asymptotic formula[28] for the upper limit: $\alpha_{max,N} (rad) = (2\pi/\sqrt{3})^{1/2} N^{-1/2}$ shows that it keeps decreasing; higher values of N are then useless for our purpose. Third column: the corresponding volume variation for the N inverted caps.

For up to N = 4, the necessary condition is not sufficient. This is obvious for N=2: the caps corresponding to the solutions of the Tammes' problem are two hemispheres (α=90°), which cannot be reverted both simultaneously. Interpenetration of two opposite depressions begins when they contact at the center of the sphere, which corresponds to a maximum value of the depression half-angle α: $\alpha_{max,2}$ = 60°. We will repeat a likewise operation in what follows: keeping the centers of the spherical caps of the Tammes' problem solutions at their positions, then decrease α down to a value for which there is no interpenetration. For N=3, the caps that solves the Tammes' problem are centered on the vertices of an equilateral equatorial triangle[24]; a little trigonometry show that interpenetration is avoided when α reduces to $\alpha_{max,3}$ with: $\cos \alpha_{max,3} = 1/\sqrt{3}$ (Fig. 4). In a similar way, one can show for N=4 that $\cos \alpha_{max,4} = 1 / (2 \sin \alpha_T/2)$, where $\alpha_T = 109°$ is the angle between two center/vertex directions in the tetrahedron. Hence the half-angle of 54.59° (Tammes' problem solution for N=4, with tetraedric symmetry) corresponds to 52.24° for the non-interpenetrating inverted caps. For higher values of N (N>4), the polyhedron holding the bases of the spherical caps present only right or obtuse angles between adjacent faces: there is therefore no risk of interpenetration as long as the caps bases do not overlap – and the Tammes' condition is sufficient.

With these values of $\alpha_{max,N}$, the maximum relative volume variation $(\Delta V/V_{sphere})_{max,N}$ can be calculated for each N (Table I). For all N, $\Delta V/V_{sphere}$ has no physical meaning above $(\Delta V/V_{sphere})_{max,N}$ (as shown before, the mathematical solution provided by equations (3a) and (3b) corresponds to interpenetration of the inverted caps beyond this limit) : hence the energy $U_N$ of the state with N depressions is plotted and considered only up to $(\Delta V/V_{sphere})_{max,N}$ in figure 3. We also restrict to values of N from 1 to 6, because higher values correspond to energies superior to $U_6$ for physical values of $\Delta V/V_{sphere}$ (*i.e.* ; smaller than $(\Delta V/V_{sphere})_{max,N}$).

Looking for the number N of depressions that corresponds to the lowest energy at every relative volume variation provides the phase diagram displayed on Fig. 5: at first, the optimum N increases with $\Delta V/V_{sphere}$ due to energetic considerations, from N=1 up to a quite stable cubic organization of the depressions (N=6). The cap half-angle α keeps a value around 40° during this evolution. Then steric factors favor a tetraedric-related conformation (N=4), followed by the final single depression state. It is interesting to note that the state N=1 for high relative volume variations is made compulsory by the geometry: this explains *a posteriori* why there is no point in refining the energy calculation in this limit.

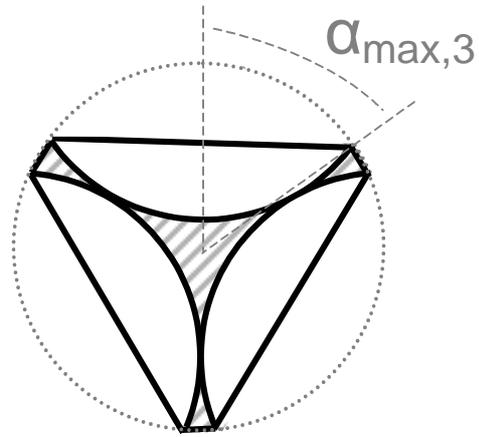

*Figure 4:* Equatorial section of the N = 3 conformation: $\alpha_{max,3} = 54,74°$. The solution to Tammes' problem would correspond to slightly larger spherical caps ( α = 60°) that could not revert without interpenetration.

In observations by Gao et al[8], which correspond obviously to a zero spontaneous curvature case as the capsule is composed of symmetrical polyelectrolyte layers, single depressions were observed for angles up to 46° for osmotic pressures of 22 kPa, and beyond 76° for 44.5 kPa (increase of osmotic pressure leads to a loss of inner volume), as displayed in Fig. 1b. This is not far from the upper and lower limits for N=1, which suggests that it could be worthwhile exploring intermediate values of the osmotic pressure in order to check the presence of configurations presenting a higher number of depressions. In a similar but thinner system (d/R = 0.018), a transition from N=1 towards a higher number of depressions was observed when increasing the osmotic pressure[25], but a further increase leads to a zoology of tortured shapes for which we do not have a model. With localized occurrence of important curvature, coalescence of distinct depressions can be prevented by high energy barriers. This would lead to quenching of the system in metastable conformations, which is not rare in buckling problems.

In the present model of zero spontaneous curvature, a single parameter (the total relative volume variation) happens to drive the conformations; this is worthwhile to be commented. As previously stated, we did not consider the extension of the circular ridge that continuously links the inverted cap to the spherical undeformed part. Taking it into account would (i) increase for each depression the volume variation by a term scaling in $\delta^2 R$, which means a correction in $N(d/R)$ for $\Delta V/V_{sphere}$ (ii) lower the maximum surface compactness of depressions, by adding an excluded corona around the caps (of width or order $\delta$), which amounts to decrease $d_{max,N}$ and then $(\Delta V/V_{sphere})_{max,N}$ for each N. If the prefactor of d/R in the volume correction varies slowly with $\alpha$, the global effect of (i) will be a displacement of the transitions towards smaller values of $\Delta V/V_{sphere}$, proportionnally to d/R. Besides, by affecting $(\Delta V/V_{sphere})_{max,N}$, (ii) will cause an additional displacement of the "steric" transitions (*i.e.* transitions towards smaller values of N). Another consequence is a concommittant lowering of the range of $\alpha$ for the corresponding conformations, decreasing in particular the minimum value of $\alpha$ that can be obtained for the capsule. Both effects vanish for the thinnest shells, which is likely to be the case in aforementioned references.

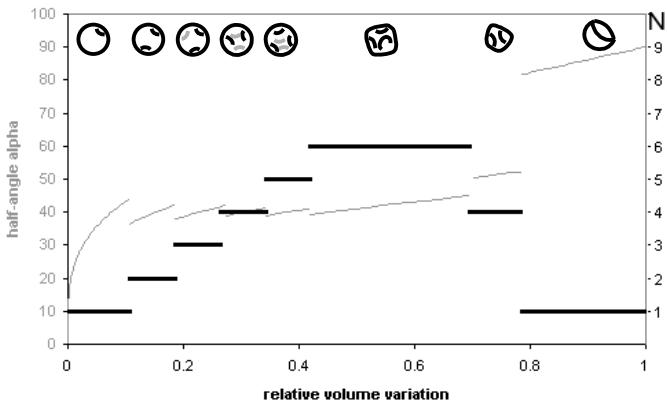

*Figure 5:* Black (right axis) : equilibrium values of the number of similar depressions N at the surface of the sphere, as a function of the relative volume variation $\Delta V/V_{sphere}$. The five first transitions occur at $\Delta V/V_{sphere} = 0.106, 0.185, 0.274, 0.341$ and $0.418$. The following "steric" transitions occur at values given in Table I for N=6 and N=4. Grey (left axis) : in degrees, half-angle $\alpha$ of the depressions.

## 3. Shells of nonzero spontaneous curvature

In this section, the reference state of the shell surface has a (algebric) spontaneous curvature $c_0$ (surface being oriented towards the outside: for example, an initially unstrained case corresponds to $c_0 = +1/R$). Curvature energy per unit surface then writes $\kappa(c-c_0)^2/2$ instead of $\kappa c^2/2$, where c is the local mean curvature.

The elastic energy of the ridge[19] can then be expressed as:

$U_{ridge} = 2\pi\kappa\,\delta R \sin\alpha\,((\delta/tg\,\alpha)^{-1} - c_0)^2$.

In the region enclosed by the ridge, changing the curvature of a spherical cap of surface $2\pi R^2 (1 - \cos\alpha)$ from $(+1/R)$ to $(-1/R)$ costs an energy:

$U_{cap} = 2\pi\kappa R^2 (1 - \cos\alpha)\,4 c_0 / R$.

The total elastic energy of a single depression then writes:

$$U_{1,\lambda} = 2\pi\kappa\,(d/R)^{-1/2}\,[\sin\alpha\,(tg\,\alpha - \lambda)^2 + 4(1 - \cos\alpha)\lambda]$$
(5)

where $\lambda = c_0 R\,(d/R)^{1/2}$.

Equation (5) shows that the relative volume variation is not anymore the only parameter that drives the conformations: the relative thickness d/R and the relative spontaneous curvature $c_0 R$ also play a qualitatively similar role.

For various values of $\lambda$, we used this $U_{1,\lambda}$ like $U_1$ in §2, in order to trace $U_{N,\lambda}$ as a function of $\Delta V/V_{sphere}$ for different numbers of depressions N, and then to determine which N has the lowest energy. Results are summarized in figure 6, that displays the phase diagram of the optimal N according to $\lambda$ and $\Delta V/V_{sphere}$. In particular, the angle of a small single depression (corresponding to the upper limit of the N=1 region for low relative volume variations) varies from 44° for $\lambda = 0$ to 60° for $\lambda = -1.2$ and to 66° for $\lambda = 1.2$.

For comparison with experimental results, we can remark that due to *in situ* formation, the elastic shell composed of aggregated polysterene colloids of reference[7] should be unstrained, hence $c_0 R=1$ and $\lambda = (d/R)^{1/2} = 0.3$ in this case. The phase diagram shows that one can find a range of $\Delta V/V_{sphere}$ presenting a N=6 state for this value.

For what concerns the *capsules* of reference[6,26], it is hard to propose an *a priori* value for $\lambda$. The templated spherical shells these capsules were obtained from must have been strainless after their synthesis, but subsequent modification of inner and outer solvant is likely to have caused a nonzero spontaneous curvature due to asymetric surface chargeing.

Interestingly enough, the steric boundaries of the phase diagram are not affected by λ, hence the observation of capsules with a single depression of half-angle ~90° is consistent with the position of the upper N=1 zone, whatever the value of λ.

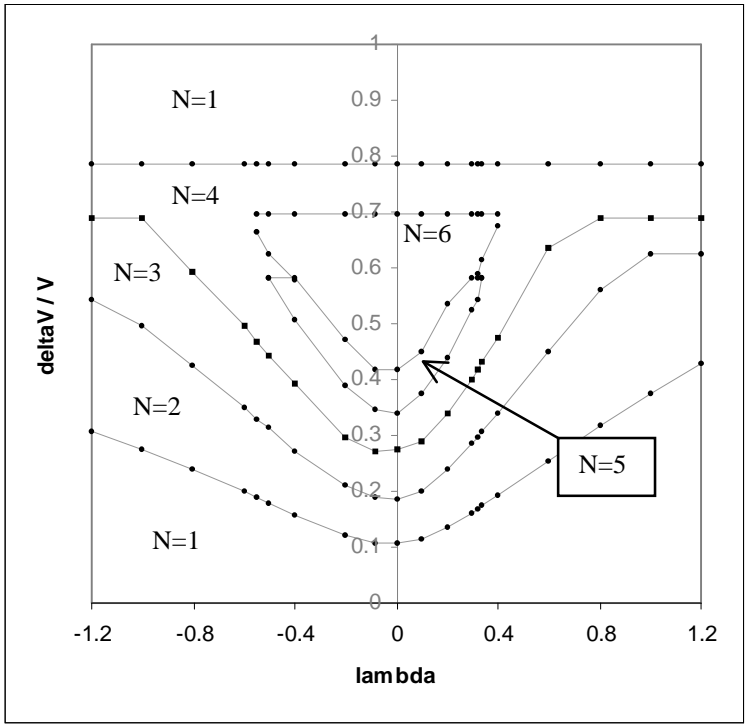

*Figure 6: Phase diagram displaying the equilibrium value of N (number of similar depressions) for shells of nonzero spontaneous curvature $c_0$. Vertical axis : relative volume variation $\Delta V/V_{sphere}$. Horizontal axis : adimensional curvature parameter $\lambda = c_0 R (d/R)^{1/2}$.*

## 4. Conclusion

We showed that a relatively simple model of elastic deformations at equilibrium, and mainly by curvature, is sufficient to make the first predictions concerning postbuckling conformations of a spherical shell submitted to an external pressure. In this model, the parameters that drive the transitions between different states (*i.e.* different number of depressions formed by inversion of a spherical cap) are the total relative volume variation and the adimensional curvature parameter λ taking into account both the thickness of the shell and its spontaneous curvature (rescaled by the initial sphere radius). The buckled sphere exhibits a single depression for both small and important relative volume variations, and several depressions (up to 6, which leads to a cubic symmetry) for intermediate ones. This model proposes a phase diagram that is likely to explain the conformations observed in experiments: "capsules" with a nearly hemispheric single depression for important variations of the inner volume, observed by Zoldesi et al[6,26], other N=1 states observed by Gao et al[8], or coarsely cubic shapes for weaker variations observed by Tsapis et al[7]. For more accurate predictions, a more sophisticated model considering the geometry of the ridge is required. It should reveal, at higher orders, an extra influence of the relative shell thickness d/R on the boundaries of the phase diagram.

Acknowledgements : The author thanks E. B. Saff, R. Ferreol and J.-P. Raven for interesting suggestions, F. Graner and J.-P. R. for careful reading of the manuscript and B. Houchmandzadeh for his introduction to Scilab. Financial support was granted from D. G. A. (Direction Générale de l'Armement) and Transregio Template Program.